\documentclass[10pt,twocolumn,citeautoscript,floatfix,aps,prb,
      superscriptaddress]{revtex4-2}

\usepackage{amsmath}
\usepackage[T1]{fontenc}
\usepackage{mathtools}
\usepackage{physics}
\usepackage{appendix}

\usepackage{graphicx}
\graphicspath{{./images/}}

\usepackage{epstopdf}
\usepackage{natbib}
\usepackage{multirow}
\usepackage{array}
\usepackage[usenames,dvipsnames]{color}
\usepackage{dcolumn}%
\usepackage{bm}%
\usepackage{amssymb} %
\usepackage{booktabs}  
\usepackage{siunitx}

\definecolor{my_blue}{RGB}{60,200,230}

\usepackage{soul} 
\usepackage{changes}

\usepackage[colorlinks=true,linkcolor=blue,citecolor=blue,urlcolor=magenta]{hyperref}

\makeatletter
\let\Hy@footnote@currentHref\relax
\makeatother

\begin{document}
\title{Piezogyration and elastogyration in ternary chalcopyrites from first principles}
\author{Asier Zabalo}
\affiliation{Physique Théorique des Matériaux, Q-MAT, Université de Liège, B-4000 Sart-Tilman, Belgium}
\author{Fernando G\'omez-Ortiz}
\affiliation{Physique Théorique des Matériaux, Q-MAT, Université de Liège, B-4000 Sart-Tilman, Belgium}
\author{Aldo H. Romero}
\affiliation{Department of Physics and Astronomy, West Virginia University, Morgantown, WV 26505-6315, USA}
\author{Eric Bousquet}
\affiliation{Physique Théorique des Matériaux, Q-MAT, Université de Liège, B-4000 Sart-Tilman, Belgium}
\date{\today}
\begin{abstract}
Natural optical activity arises from the spatial dispersion of the optical response and is described by the rank-two gyration pseudotensor.
Its dependence on mechanical stress and strain, quantified by the piezogyration and elastogyration tensors, respectively, has been investigated experimentally for decades. However, a systematic first-principles description of these effects remains largely unexplored.
Here we present a first-principles methodology to compute the full piezogyration and elastogyration tensors,
and apply it to a set of representative ternary chalcopyrite crystals.
The gyration of the unstrained crystal turns out to be a poor predictor of the magnitude of the elastogyration response, which we find to be a property inherited from their
parent zinc-blende structures. 
The elastic compliance tensor further shapes the piezogyration response, with particularly pronounced effects in AgGaS$_2$, a well-known achiral yet optically active crystal.
Finally, we explore
strain-induced
chirality through the trace of the gyration tensor,
the only pseudoscalar contribution to the natural optical activity 
that is linear in the tensor components and
encodes information about chirality.
\end{abstract}

\maketitle

\section{Introduction}
Ternary chalcopyrite compounds of stoichiometry ABC$_2$, with space group $I\bar{4}2d$, constitute one of the most 
versatile families of tetrahedrally-coordinated
semiconductors known to date~\cite{shay2013ternary}. Structurally, the chalcopyrite unit cell can be viewed as a body-centered tetragonal 
superlattice built from two stacked zinc-blende cells, AC and BC, in which the two cation sublattices are ordered along the
crystallographic c axis. 
Despite their apparently simple crystal structure, ternary 
ABC$_2$ chalcopyrite crystals continue to attract considerable attention both for their potential technological applications~\cite{Siebentritt2022,10.1063/1.1655537} and as platforms for exploring fundamental phenomena in condensed matter physics, including topologically
non-trivial phases~\cite{PhysRevB.83.245202,PhysRevLett.106.016402}, Weyl semimetals~\cite{PhysRevLett.116.226801,qdvf-q7w9}, pressure-induced topological phase transitions~\cite{PhysRevB.106.125112},
and non-linear optics~\cite{PhysRevMaterials.4.064602}.

In this work, we examine the piezogyration properties of ternary chalcopyrite crystals, an optical response that has attracted relatively little attention. Piezogyration is defined as the stress derivative of the gyration tensor, which characterizes natural optical activity (NOA).
K. Aizu first formulated this response in his 1964 symmetry analysis of ferroelectric tensor properties~\cite{PhysRev.133.A1350}; the equivalent formulation in terms of strain is usually referred to as \textit{elastogyration}~\cite{Martynyuk-Lototska2010,Weber1979}.
Building on this idea, in the late 1960s, G. S. Ranganath 
and S. Ramaseshan~\cite{Ranganath1969JOSA,Ranganath1969PASA} developed a phenomenological theory of strain-mediated optical activity and piezo-rotatory coefficients, valid at first-order in the applied deformation field. 
Since the early symmetry foundations and the first experimental works,  
strain-induced gyrotropic effects in crystals, 
both uniform and nonuniform, have attracted
renewed interest as a route to engineer, enhance, and induce optical 
activity~\cite{Vasylkiv2013,skab2011optical,vasylkiv2013optical,Vlokh1997,
vlokh1989symmetry,mys2007piezogyration,
10.1039/c9ra03696b,zk45-6lb2,
Konak_1978,ETCHEGOIN1992655}.

However, even with the current experimental advancements, extracting all the piezogyration coefficients of a crystal remains a formidable challenge.
Measuring the natural optical activity of crystals along
directions other than the optic axis is difficult,  
since the much larger linear birefringence tends to mask the effect~\cite{Kaminsky-00}.
The difficulty is further amplified when the optical
activity is strain-induced: from a symmetry standpoint,
the only way mechanical deformations can generate new NOA tensor elements is when the applied strain—whether uniform or nonuniform—reduces the crystal symmetry, causing the system to transition into a different point group. In such situations, the resulting optic
axis, whether newly created by the strain or simply
reoriented from its unstrained position, is generically
not aligned with any crystallographic axis but instead
depends on the direction and magnitude of the applied
deformation field.

Here, we show that first-principles computations using density-functional perturbation theory (DFPT) enable direct determination of all piezogyration tensor coefficients efficiently and accurately.
We focus exclusively on the ternary
chalcopyrite 
crystals,~\cite{AUSTIN1956}
a family of highly tunable direct band-gap semiconductors
where strain-induced gyrotropic effects
are yet to be fully explored. 
We select eight representative chalcopyrite crystals spanning 
the two major structural and chemical families of this class of materials. Specifically,
AgGaS$_2$ and CuAlSe$_2$, belonging to the I-III-VI$_2$ family, and 
the two groups 
MgSiP$_2$, MgGeP$_2$, and MgSnP$_2$ and
ZnSiP$_2$, ZnGeP$_2$, and ZnSnP$_2$, belonging to the II-IV-V$_2$ family.

The rest of the manuscript is organized as follows. In Sec.~\ref{sec_theory}, we 
summarize the symmetry restrictions on the natural optical activity and piezogyration tensor 
of ternary ABC$_2$ chalcopyrite crystals with  $\bar{4}2m$ point group symmetry.
In Sec.~\ref{sec_results}, we present our first-principles results of piezogyration
in the selected structures and discuss the 
structural and electronic effects that govern the optical response.
In addition to examining the piezogyration response, we also
compute the elastogyration coefficients, formulated in terms of strain
rather than stress. To the best of our knowledge, this work is the first to compute the full piezogyration and elastogyration tensors for an entire materials family using first-principles calculations.
Our results indicate that the elastogyration tensor of the considered ternary 
chalcopyrite crystals are largely inherited from their parent zinc-blende
structures, whose $\bar{4}3m$ point group allows only a single independent
elastogyration coefficient. 

We focus to a great extent on the $R_{31}$ elastogyration coefficient  of the chalcopyrites, which triggers a response that is forbidden by symmetry in the unstrained phase, 
namely optical rotation for light propagating along the $c$ axis. 
The sign of this coefficient is the same in all the compounds.
We account for this observation by considering the orientation of the cation–anion tetrahedra in the chalcopyrites, along with the direction of bond polarity, which is identical across all the crystals examined here.
In Sec.~\ref{Sec_elastochiral}, we briefly discuss the related piezochiral (elastochiral) effect, 
namely, the emergence of chirality induced by mechanical stress (strain), 
anticipated in Ref.~\cite{bousquet2025structural} and 
recently formalized and inferred experimentally in AgGaS$_2$ from the measured optical rotation along the crystal's $c$-axis~\cite{Zeng2026}.
Our calculated value for the strain-induced optical rotation in silver gallium sulfide agrees well with experiments.
Finally, Sec.~\ref{sec_conclusions} summarizes our main findings and conclusions.

\section{Theory}\label{sec_theory}
In this section, we summarize the theoretical framework
underlying natural optical activity and piezogyration,
focusing on the definitions and tensorial relations
needed for our first-principles calculations.
For a broader discussion of these phenomena,
we refer the reader to
Refs.~\cite{malgrange2014symmetry,barron2009molecular,
nye1985physical,agranovich2013crystal,
sirotin1982fundamentals}.

\subsection{Optical rotatory power and gyration tensor}
Natural optical activity arises due to the dependence of the dielectric tensor,
$\epsilon_{ij}$,
on the 
wave vector $\mathbf{q}$, and can be conceived as a first-order spatial dispersion effect. Mathematically,
\begin{equation}
    \epsilon_{ij}(\omega,\mathbf{q})\simeq \epsilon_{ij}(\omega,\mathbf{q=0})+iq_k\eta_{ijk}(\omega),
\end{equation}
where $\eta_{ijk}$ is responsible for the NOA and implicit summation of repeated subscripts is assumed
throughout the whole text~\footnote{Similarly, it should be understood that the dielectric tensor $\epsilon_{ij}$ refers to the 
electronic part of the dielectric permittivity tensor, commonly indicated with the $\epsilon_{ij}^\infty$ symbol.}.
In the transparent regime, i.e., in the absence of dissipation, the dielectric tensor is Hermitian. It follows that $\eta_{ijk}(\omega)=-\eta_{jik}(\omega)$, 
indicating antisymmetry in its first two indices and an intrinsic symmetry type
$\{V^2\}V$. 
Throughout this work, we consider the well-defined zero-frequency limit of natural optical activity. We therefore omit the explicit frequency dependence $\omega$ from the notation.
It is convenient to introduce the second-rank pseudotensor of gyration $\boldsymbol{g}$, satisfying
\begin{equation}
    g_{ij}=\frac{1}{2}\varepsilon_{nmi}\eta_{nmj},
\end{equation}
where $\varepsilon_{nmi}$ is the Levi-Civita symbol. Similarly, the dual transformation is obtained as
$\eta_{ijk}=\varepsilon_{ijm}g_{mk}$. 
The constitutive equation 
relating the electric displacement $\mathbf{D}$ with the electric field $\mathbf{E}$
in terms of the gyration pseudotensor is therefore given  by
\begin{equation}\label{Eq_D}
D_i=\left(\epsilon_{ij}+i\varepsilon_{ijm}g_{ml}q_l\right)E_j. 
\end{equation}
The main question is, of course, finding out how the spatial dispersion term that is characterized by the 
second-rank pseudotensor $\boldsymbol{g}$
affects wave propagation in crystals, and more importantly, observable experimental effects.
In particular, the main physical manifestation of NOA is the optical rotation, i.e., 
the rotation of the plane of polarization of linearly polarized light
as it passes through the crystal. 
For an arbitrary propagation direction, the Fresnel equation, including the spatial dispersion introduced through the constitutive relation in Eq.~(\ref{Eq_D}), generally yields elliptically polarized eigenmodes. Their polarization depends on the interplay between linear birefringence and circular birefringence associated with natural optical activity. As a result, the evolution of the polarization cannot, in general, be described as a simple optical rotation. The situation becomes simpler along an optic axis, where linear birefringence vanishes by symmetry. In this case, the polarization eigenmodes are circularly polarized, and the optical rotation, $\rho(\omega)$, is determined entirely by natural optical activity.
The zero-frequency limit of this quantity is usually 
indicated with the symbol $\bar{\rho}$, and is given by~\cite{Ramachandran1951}
\begin{equation}\label{Eq_rho}
   \bar{\rho} =\lim_{\omega\to 0}\frac{\rho(\omega)}{(\hbar\omega)^2} = \frac{G}{2(\hbar c)^2}.
\end{equation}
Here, $c$ is the speed of light, $\hbar$ the reduced Planck constant, and $G=l_ig_{ij}l_j$, where
$l_i$ are the direction
cosines of the wave vector $\mathbf{q}$. This result highlights that only the symmetric part 
of $g_{ij}$ contributes to 
optical rotation. 
It is then convenient to write 
\begin{equation}
    g_{ij}=g_{(ij)}+g_{[ij]},
\end{equation}
where
\begin{equation}\label{Eq_sym_antisym}
    \begin{split}
        g_{(ij)}&=\frac{1}{2}\left(g_{ij}+g_{ji}\right),\\
        g_{[ij]}&=\frac{1}{2}\left(g_{ij}-g_{ji}\right).
    \end{split}
\end{equation}
The symmetric tensor $g_{(ij)}$ is more conveniently represented 
in Voigt notation by the six-component gyration vector 
$g_I$ ($I=1,...,6$). Explicitly, 
\begin{equation}
    g_I=(g_{(11)},g_{(22)},g_{(33)},g_{(23)},g_{(13)},g_{(12)}).
\end{equation}
From now on, we shall focus our analysis on this quantity.
\subsubsection{$\bar{4}2m$ point group crystals}\label{sec_point_group}
Symmetry requirements dictate that, out of the 32 crystallographic point groups,
NOA is present in 18 of the 21 acentric crystal classes, including all 11 enantiomorphic groups.
Note, however, that three of the 18 optically active crystal  
classes, namely 3m, 4mm, and 6mm, do not allow optical rotation as the
gyration tensor is purely antisymmetric~\cite{10.1063/1.433207}.
The only three acentric crystal classes that do not exhibit optical activity are $\bar{6}$, $\bar{6}2m$, and $\bar{4}3m$~\cite{barron2009molecular,Urru2025}.
Although optical activity has historically been associated primarily with chiral crystals,
it soon became evident that chirality  
is not a necessary condition for natural 
optical activity.
In fact, the ternary chalcopyrite AgGaS$_2$ is the first experimentally verified and most iconic example of an optically active yet achiral crystal~\cite{hobden1968optical}. 
In particular, crystals belonging to space groups compatible with 
$\bar{4}2m$ point group symmetry, among which the ternary chalcopyrites  belong,
exhibit the following gyration tensor,
\begin{equation}\label{Eq_g_vec}
g=\begin{pmatrix}
    g_1&0&0\\
    0&-g_1&0\\
    0&0&0
\end{pmatrix}\longrightarrow 
    g_I=(g_1,-g_1,0,0,0,0),
\end{equation}
where there is only one independent component.
The trace of the gyration tensor, which transforms as a pseudoscalar,
is defined as $\Tr(g)=g_1+g_2+g_3$. From Eq.~(\ref{Eq_g_vec}), we see that this quantity is trivially zero
for $\bar{4}2m$ point group crystals, consistent with optically active yet achiral 
systems~\cite{10.1063/1.433207}. $\Tr(g)$ is a pseudoscalar under every rotation symmetry (i.e., invariant under proper rotations and odd under improper ones), and it vanishes in any point group
containing improper symmetry operations, regardless of the reference structure or 
frame.~\footnote{The determinant of the gyration tensor, $\det(g)$, 
is also a pseudoscalar and vanishes in achiral
systems. However, it is cubic in the components $g_{ij}$, so
$\Tr(g)$ is the only pseudoscalar invariant linear in $g$.}
This means that 
$\Tr(g)\neq 0$ only if the system is chiral, though accidental zeros in chiral structures 
may occur~\cite{pelloni2017electric}.
\subsection{Piezogyration and elastogyration in $\bar{4}2m$ crystals}\label{Sec_elasto_theory}
The piezogyration effect is defined as the change of the gyration vector $g_I$ at first-order
in the applied uniform stress field,
\begin{equation}
	\Delta g_I=\Pi_{IJ}\sigma_J,
\end{equation}
where $\Pi_{IJ}$ ($I,J=1,...,6$) is the piezogyration tensor and $\sigma_J$ represents the stress in Voigt notation~\footnote{The internal ionic coordinates are relaxed at each value of the deformation field, meaning
that all the coefficients reported in this work are relaxed-ion quantities.}.
Equivalently, the changes in the gyration vector could be expressed as linear in the applied strain, such that
\begin{equation}\label{Eq_delta_g_R}
	\Delta g_I=R_{IJ}s_J,
\end{equation}
where $R_{IJ}$ are the elastogyration coefficients and $s_J$ represents the strain.
The two descriptions are, of course, equivalent and carry the same information; they are related through
\begin{equation}\label{Eq_T_from_R}
	\begin{split}
	R_{IJ}&=\Pi_{IK}\mathcal{C}_{KJ},\\
	\Pi_{IJ}&=R_{IK}\mathcal{S}_{KJ},
	\end{split}
\end{equation}
where $\mathcal{C}_{IJ}$ and $\mathcal{S}_{IJ}$ are, respectively, the (relaxed-ion) elastic 
stiffness and compliance tensors.~\footnote{Note that different authors have used distinct symbols for the elastogyration and piezogyration coefficients. For example,  in 
Refs.~\cite{Ranganath1969JOSA,Ranganath1969PASA} 
Ranganath and Ramaseshan use the symbols $S$ and $R$ to denote strain and stress, respectively, which differs from the convention adopted here.}~

Non-vanishing components of both the piezogyration and elastogyration tensors for all 32 crystallographic 
point groups are dictated by group theory analysis~\cite{Aroyo2011}. 
Focusing on crystals with point group $\bar{4}2m$,
we find that symmetry allows five independent piezogyration coefficients~\cite{Ranganath1969JOSA},
namely $\Pi_{11}$, $\Pi_{21}$, $\Pi_{31}$, $\Pi_{55}$ and $\Pi_{13}$. In matrix form:
\begin{equation}\label{Eq_R_matrix}
\Pi_{IJ}=
\begin{pmatrix*}[r]
\Pi_{11} & -\Pi_{21} & \Pi_{13} & 0 & 0 & 0 \\
\Pi_{21} & -\Pi_{11} & -\Pi_{13} & 0 & 0 & 0 \\
\Pi_{31} & -\Pi_{31} & 0 & 0 & 0 & 0 \\
0 & 0 & 0 & -\Pi_{55} & 0 & 0 \\
0 & 0 & 0 & 0 & \Pi_{55} & 0 \\
0 & 0 & 0 & 0 & 0 & 0
\end{pmatrix*}.
\end{equation}
The piezogyration coefficients are evaluated as
\begin{equation}\label{Eq_piezo_def}
	\Pi_{IJ}=\frac{\partial \Delta g_I}{\partial \sigma_J}\Big|_{\boldsymbol{\sigma}=0},
\end{equation}
which complies with the definition given by Aizu in 1964: \text{\guillemotleft}The ``piezogyration'' tensor means
the tensor equal to the rate of change of the gyration tensor with the stress tensor at zero value of the 
stress tensor.\text{\guillemotright}~\cite{PhysRev.133.A1350}. 
Similarly, the elastogyration coefficients are extracted from Eq.~(\ref{Eq_delta_g_R}) as the following partial derivatives 
evaluated at the equilibrium state, with all remaining strain components held fixed at zero:
\begin{equation}\label{Eq_elasto}
	R_{IJ}=\frac{\partial \Delta g_I}{\partial s_J}\Big|_{\mathbf{s}=0}.
\end{equation}

In the following, we shall consider the strain as our external perturbing field
and restrict our discussion to the elastogyration tensor.
This analysis trivially generalizes 
to the case in which stress is the main driving force of the effect.
While some elastogyration coefficients modulate already existing gyration vector entries
---this is the case of $R_{11}$, $R_{21}$ and $R_{13}$---, others represent
symmetry-breaking couplings that enable new, strain-induced gyration components, namely
$R_{31}$ and $R_{55}$.
This distinction is made explicit below
for different strains, where the induced changes in the gyration tensor can be directly compared to
Eq.~(\ref{Eq_g_vec}), which shows the shape of $g$ in the unstrained phase.

Uniaxial strain along $c$, denoted by $s_3$, preserves every symmetry element of the
$\bar42m$ point group, particularly the improper $S_4$ axis along $c$, and
therefore cannot induce any new gyration-tensor component beyond $g_1=-g_2$,
which is already present in the unstrained crystal. We employ the 
symbol $\mathcal{F}$ to indicate the deformation matrix acting on the lattice vectors
induced by the applied strain, which specifies the \textit{complete} strain state with the cell held clamped, consistent with 
Eq.~(\ref{Eq_elasto}):
\begin{equation}
	\mathcal{F}=\begin{pmatrix} 1&0&0\\ 0&1&0\\ 0&0&1+s_3 \end{pmatrix}
	\longrightarrow
	\Delta g = \begin{pmatrix} \Delta g_1&0&0\\ 0&-\Delta g_1&0\\ 0&0&0 \end{pmatrix}.
\end{equation}

Uniaxial strain along $a$, denoted by $s_1$,  plays a qualitatively different role, as it 
breaks the $S_4$ axis and the two diagonal
mirror planes, lowering the point group from $\bar42m$ ($D_{2d}$) to $222$
($D_2$).
In the absence of the $S_4$ symmetry operation, the gyration component $g_{3}$, 
which is forced to vanish in 
the undistorted $\bar{4}2m$ phase, 
is no longer forbidden by symmetry. 
This turns on an entirely new optical response: optical rotation for light propagating parallel to the $c$ axis.
Explicitly,
\begin{equation}
\mathcal{F}=\begin{pmatrix} 1+s_1&0&0\\ 0&1&0\\ 0&0&1 \end{pmatrix}
\longrightarrow
\Delta g=
    \begin{pmatrix}
	         \Delta g_1&0&0\\
	        0& \Delta g_2&0\\
	        0&0&\Delta g_3\\
	    \end{pmatrix}.
\end{equation}
Note that $\Tr(\Delta g)\neq 0$, indicating that $s_1$ induces chirality at first-order in the applied strain 
field~\cite{10.1063/1.433207}.
Strictly speaking, 
and as noted in Ref.~\cite{Zeng2026}, the chiral coupling transforms as $s_{xx}-s_{yy}$. 
A purely uniaxial strain $s_{xx}$, however, 
can be trivially decomposed as 
$s_{xx} = \frac{1}{2}(s_{xx}-s_{yy})+\frac{1}{2}(s_{xx}+s_{yy})$, i.e., into a chiral and a non-chiral component; 
the chiral term alone is sufficient to lower the symmetry. 
Induced chirality via mechanical deformation has been previously described 
as either piezochirality~\cite{bousquet2025structural,Zeng2026}
or elastochirality~\cite{bousquet2025structural} 
when formulated in terms of stress and strain, respectively.
The elastochiral (piezochiral) effect can be regarded a sub-effect of the more general 
elastogyration (piezogyration) effect~\cite{Konak_1978}. 
The latter is present whenever uniform strain (stress) modifies the gyration vector $g$, while the former focuses solely on the trace of $g$. We will come back to this point
in Sec.~\ref{Sec_elastochiral}.

Shear strain $s_5$ lowers the point group further still, from
$\bar42m$ to $2$.~\footnote{Note that even if the crystal becomes chiral, 
$\Tr(g)$ remains zero at first order in the applied strain field, though higher order coupling
terms do produce  $\Tr(g)\neq 0$. Throughout this work, we shall be concerned with strain and stress effects at first order only.}
Here it is
the off-diagonal component $g_5\equiv g_{(13)}$ that is switched on, while $g_1$ and
$g_2$ remain unchanged at linear order in strain:
\begin{equation}
	\mathcal{F}=\begin{pmatrix} 
		1&0&s_5/2\\ 
		0&1&0\\ 
		s_5/2&0&1 \end{pmatrix}
	\longrightarrow
\Delta g=
    \begin{pmatrix}
	         0&0&\Delta g_5\\
	        0&0&0\\
	        \Delta g_5&0&0\\
	    \end{pmatrix},
\end{equation}
with the convention $s_5=2s_{13}$.

\section{Results}\label{sec_results}
In this section, we show the results for the 
ternary ABC$_2$ chalcopyrite crystals under consideration:
AgGaS$_2$, CuAlSe$_2$, 
MgSiP$_2$, MgGeP$_2$, MgSnP$_2$,
ZnSiP$_2$, ZnGeP$_2$, and ZnSnP$_2$.
\subsection{Computational parameters}
We perform our first-principles calculations with DFT and DFPT as implemented
in the \textsc{abinit}~\cite{Verstraete2025,Romero2020} package (v9.9.3).
We use norm-conserving scalar relativistic pseudopotentials under the PBEsol parametrization
for the exchange-correlation functional from the Psuedodojo~\cite{vanSetten2018} website (v0.4.1),
with a plane-wave energy cutoff of $40$ Ha in all cases.
We employ a $12\times 12\times 6$ $\mathbf{k}$-point mesh for all materials, 
except for ZnSnP$_2$, for which a denser  $14\times 14\times 7$  $\mathbf{k}$-point
mesh is used.
Structures are relaxed to mechanical equilibrium until forces and stresses are smaller than 
$10^{-7}$ Ha/Bohr and $10^{-9}$ Ha/Bohr$^3$, respectively.

We compute the natural optical activity using the
methodology described in Ref.~\cite{Zabalo2023}, as implemented in 
\textsc{abinit}~\cite{Verstraete2025,Zabalo2023,s55b-g2vt}. We do not include spin-orbit coupling in our calculations.

Appendix~\ref{App_strategy} provides our computational strategy for the calculation of 
both the piezogyration and elastogyration
coefficients, where the derivatives with respect to stress and strain are taken via 
central finite differences.

\subsection{Structural, electronic and optical properties of the unstrained phase}\label{Sec_strc}

\begin{figure}[t]
	\includegraphics[width=\columnwidth]{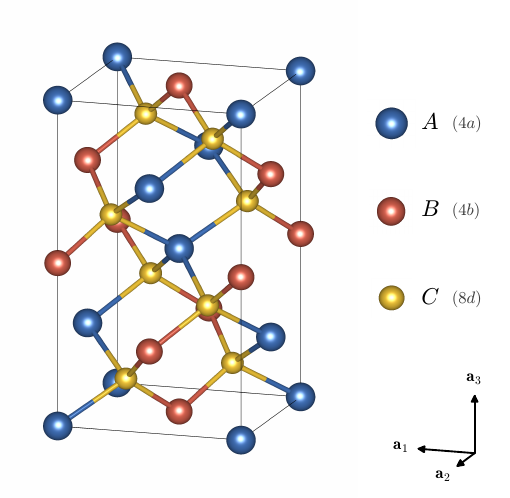}
	\caption{Cartoon illustrating the crystal structure of a generic ternary chalcopyrite 
    crystal with stoichiometry ABC$_2$.}
	\label{fig_ABC2_strc}
\end{figure}

We provide in  Table~\ref{Tab_wyck} the Wyckoff positions for the ternary ABC$_2$ chalcopyrite crystals, where the cations A and B occupy the 
$4a$ and $4b$ sites, respectively, and the anion C occupies the $8d$ site. 
That is, A$=\{$Ag, Cu, Zn, Mg$\}$,
B$=\{$Ga, Al, Si, Ge, Sn$\}$ and C$=\{$S, Se, P$\}$  for the compounds considered here.
A schematic representation of the crystal structure is given in Fig.~\ref{fig_ABC2_strc}.
\begin{table}[t!]
	\caption{Atomic positions of the ternary chalcopyrite structure $AB$C$_2$
		(with space group $I\bar{4}2d$) in the conventional body-centered tetragonal
		cell containing 16 atoms, whose lattice vectors lie along the Cartesian axes,
		$\mathbf{a}_1=a\,\hat{\mathbf{x}}$, $\mathbf{a}_2=a\,\hat{\mathbf{y}}$,
		$\mathbf{a}_3=c\,\hat{\mathbf{z}}$. Every site is to be taken together 
		with the translation
		$(\frac{1}{2},\frac{1}{2},\frac{1}{2})$. The tetragonal distortion is 
        given by $\eta=c/2a$, and there is single free internal parameter, $u$. The ideal zinc-blende-like structure is recovered with
		$\eta=1$ and $u=\frac{1}{4}$.}
	\label{Tab_wyck}
	\begin{ruledtabular}
		\begin{tabular}{lcl}
			Atom&Wyckoff&Reduced coordinates\\ \hline \noalign{\vskip 2pt}
			$A$&$4a$&$(0,0,0)$,\quad $(0,\tfrac12,\tfrac14)$\\[4pt]
			$B$&$4b$&$(0,0,\tfrac12)$,\quad $(0,\tfrac12,\tfrac34)$\\[4pt]
			\multirow{2}{*}{C}&\multirow{2}{*}{$8d$}&$(u,\tfrac14,\tfrac18)$,\quad $(1-u,\tfrac34,\tfrac18)$,\\[2pt]
			&&$(\tfrac14,\tfrac12+u,\tfrac38)$,\quad $(\tfrac34,\tfrac12-u,\tfrac38)$
		\end{tabular}
	\end{ruledtabular}
\end{table}
The crystal 
structures of these ternary chalcopyrite crystals are closely related to
those of their binary zinc-blende analogs, but the ratio 
$\eta=c/2a$ (where $a=b$ and $c$ are lattice parameters) and the internal structural parameter $u$
deviate from their ideal values of $1$ and $1/4$, respectively. The only free structural parameters are, in fact, 
$\eta$ and $u$, with which the cation-anion bond lengths are obtained
via the following expressions~\cite{PhysRevB.28.5822},
\begin{equation}\label{Eq_bonds}
	\begin{split}
		d_{A\mathrm{C}}&=a\left[u^2+\frac{1+\eta^2}{16}\right]^{1/2},\\
		d_{B\mathrm{C}}&=a\left[\left(u-\frac{1}{2}\right)^2+\frac{1+\eta^2}{16}\right]^{1/2},
	\end{split}
\end{equation}
so that $u>1/4$ means that the $4a$ cation carries the longer bond.
The setting adopted in Table~\ref{Tab_wyck} fixes the sign 
of the gyration and elastogyration responses reported in what follows; the mirror image of the crystals described by Table~\ref{Tab_wyck} would exhibit the same gyration tensor components in magnitude, but with the opposite sign.

We provide in Table~\ref{Tab_strc} the relaxed structural parameters of the materials under study;
we collect experimental values in Appendix~\ref{app_exp}.
\begin{table}[b!]
	\caption{Lattice parameters $a=b$ and $c$ (in bohr), the ratio $\eta=c/2a$ and the internal structural parameter $u$. Values refer to the setting of Table~\ref{Tab_wyck}.  
    Experimental data can be found in Appendix~\ref{app_exp}}
	\label{Tab_strc}
	\begin{ruledtabular}  
		\begin{tabular}{lrrrr}
			&$a$&$c$&$\eta$&$u$\\\hline 
			AgGaS$_2$&10.684&19.845&0.929&0.282\\
			CuAlSe$_2$ &10.493&20.842&0.993& 0.247\\ [4pt]
			ZnSiP$_2$ &10.111&19.732&0.976&0.264\\
			ZnGeP$_2$ &10.246&20.233&0.987&0.251\\
			ZnSnP$_2$ &10.624&21.325&1.004&0.225\\[4pt]
			MgSiP$_2$ &10.837&19.032&0.878&0.293\\
			MgGeP$_2$ &10.901&19.929&0.914&0.280\\
			MgSnP$_2$ &11.132&21.663&0.973&0.253
		\end{tabular}
	\end{ruledtabular}
\end{table}
We note large variations in both the tetragonal distortion controlled by $\eta$ 
and the internal parameter $u$ across the different crystals.
Such deviations of the structural parameters from their ideal $\eta=1$ and $u=\frac{1}{4}$ values constitute
one of the ingredients of the ``band-gap anomaly'', namely the 
observation that the band-gap of these 
chalcopyrite compounds differ by more than $50\%$ 
from that of their binary zinc-blende analogs. This phenomenon was extensively studied, 
among others, by Jaffe and Zunger~\cite{PhysRevB.28.5822,PhysRevB.27.5176,PhysRevB.29.1882} already 
in the 1980s. 

Table~\ref{Tab_elec} collects the electronic properties, including 
the direct electronic band-gap at $\Gamma$ (denoted by $E_g$), the static (electronic) dielectric constants,
and the gyration coefficient $g_1$. We also show the optical rotatory power for 
light traveling parallel to the $\mathbf{a}=a\,\mathbf{\hat{x}}$
crystallographic axis, obtained by substituting $G=g_1$ in Eq.~(\ref{Eq_rho}). 
\begin{table}[b!]
	\caption{Direct electronic band-gap at $\Gamma$ ($E_g$, in eV), the independent components 
		of the electronic dielectric tensor and the only independent component of the gyration vector, $g_1$ (in bohr). The corresponding optical rotatory power $\bar{\rho}$ 
		(in deg/[mm (eV)$^2$] units) for light traveling parallel to the $\mathbf{a}=a\,\mathbf{\hat{x}}$ crystallographic axis is also reported.
		}
	\label{Tab_elec}
	\begin{ruledtabular}  
		\begin{tabular}{lrrrrr}
			&$E_g$&$\epsilon_{11}$&$\epsilon_{33}$&$g_1$&$\bar{\rho}$\\\hline 
			AgGaS$_2$&0.926&7.681&7.501&$-0.858$&$-33.4$\\
			CuAlSe$_2$ &0.922&8.426&8.432&$-0.335$&$-13.1$\\[4pt]
			ZnSiP$_2$ &1.157&10.809&10.944&$-0.579$&$-22.6$\\
			ZnGeP$_2$ &1.121&12.230&12.537&$-0.257$&$-10.0$\\
			ZnSnP$_2$ &0.789&12.641&12.592&0.214&8.3\\[4pt]
			MgSiP$_2$&1.147&8.802&9.080&0.011&0.4\\
			MgGeP$_2$&1.297&9.555&10.225&$-0.023$&$-0.9$\\
			MgSnP$_2$&1.243&9.261&9.945&0.172&6.7\\
		\end{tabular}
	\end{ruledtabular}
\end{table}

Comparing our calculated optical rotatory power in the zero-frequency limit, $\bar{\rho}$, with experimental measurements requires some care. Two factors are particularly important: (i) the differences between the calculated and experimental structural and electronic properties, and (ii) the frequency dependence of the optical rotatory power.

First, all our calculations are performed employing the predicted structure with DFT
calculations under the PBEsol functional, which might differ from the experimental one;
it has previously been 
shown that structure alone can play a significant role in determining the magnitude 
of the NOA~\cite{Zabalo2023}.
Let's take AgGaS$_2$ as a representative example.
The experimental tetragonal distortion $\eta\sim 0.895$ reported 
in Ref.~\cite{10.1063/1.1680242} shows substantial
deviation from our predicted value of $\eta=0.929$ reported in Table~\ref{Tab_strc}, and the calculated 
electronic band-gap of $0.926$ eV considerably underestimates the experimental value of
$\sim 2.7$ eV~\cite{MARCEDDU2006300,RICCI20051950}. 
Note, however, that our calculations show good agreement with previous DFT predictions; for example, 
Ref.~\cite{CHAHED2005142} reports a band-gap of $0.95$ eV computed under the local-density approximation (LDA), 
a tetragonal distortion of $\eta=0.9315$ and an internal parameter 
of $u=0.282$, essentially matching our structural values reported in Table~\ref{Tab_strc}. 
To assess the impact of structural geometry on the computed $\bar{\rho}$, 
we take AgGaS$_2$ as a test case and 	repeat our calculations with the experimental geometry 
at room temperature ($298$ K) as reported 
in Ref.~\cite{10.1063/1.1680242},  
with $a=10.880$ bohr, $c=19.471$ Bohr, and $u=0.291$.
Within this setting, the direct band-gap at $\Gamma$ is $E_g=1.037$ eV, and we obtain an optical rotatory power of 
$\bar{\rho}=-37.9$ deg/[mm (eV)$^2$] for light traveling parallel to the $\mathbf{a}=a\,\mathbf{\hat{x}}$ crystallographic axis, which is only marginally larger in magnitude than the value reported in 
Table~\ref{Tab_elec}, obtained with the predicted PBEsol structure.

Second, the optical rotatory power, $\rho(\omega)$, 
is typically probed at distinct light frequencies among different crystals. 
Note that $\rho(\omega)$ is generally strongly dispersive in frequency and 
grows rapidly (formally diverging) in the vicinity of the electronic resonances. The value $\rho(\omega)$
measured for a given 
photon energy therefore reflects not only the intrinsic strength of the rotatory power of that material,
but also how close this particular
photon energy $\hbar\omega$ is with respect to the absorption edge in that crystal. 
For example, 
the first experimental verification of optical activity in a non-enantiomorphous (i.e., achiral)
crystal was obtained by M. V. Hobden~\cite{Hobden1967,hobden1968optical} by probing
the crystal AgGaS$_2$  with a light wavelength of $\lambda=4974$ \AA, 
which corresponds to $\hbar\omega\sim 2.49$ eV (recall that the 
experimental band-gap is $\sim 2.7$ eV, and therefore, the experiment is close to measuring a divergence on the NOA
due to electronic transitions).  
The light frequency at which the crystal was probed is not 
arbitrary: Hobden realized that at such frequency AgGaS$_2$ becomes accidentally optically isotropic, 
meaning that birefringence vanishes and the measured optical rotation comes exclusively from the NOA, 
facilitating the experimental measurements. Under such conditions, Hobden reported an optical rotatory power of 
$\rho= 522$ deg/mm. From our Eq.~(\ref{Eq_rho}), we estimate $\bar{\rho}\sim 84$ deg/[mm (eV)$^2$]. 
We attribute the disagreement 
between the latter and our 
calculated value of $\bar{\rho}$---both with the PBEsol relaxed and experimental geometries---to the 
near-resonant condition at which the gyration vector component $g_1$ was probed; that is, 
$g_1(\hbar\omega\sim E_g)>g_1(\hbar\omega\to 0)$, where $E_g$ is the electronic band-gap at $\Gamma$.

Turning back to  Tables~\ref{Tab_strc} and~\ref{Tab_elec}, the first thing we note is that the magnitude of
$\bar{\rho}$ is neither govern by structural nor electronic properties alone, and further illustrates two
well-known facts: (i) a large dielectric response does not necessarily imply a large NOA 
and (ii) symmetry alone does not impose any restriction on the sign of the natural
optical activity~\cite{Glazer1986,Gomez-Ortiz:har5002}. 
The crystal AgGaS$_2$ exhibits the larger $\bar{\rho}$ in magnitude 
among the considered cases, and the Mg$B$P$_2$ family (with $B=\{$Si,Ge,Sn$\}$) shows only a marginal NOA in the unstrained phase.

The comparison between AgGaS$_2$ and MgGeP$_2$ is particularly illustrative: 
their structural distortions from the ideal zinc-blende geometry are nearly 
the same ($\eta= 0.929$ vs 0.914 and $u = 0.282$ vs 0.280), 
yet $g_1$ differs by a factor of about 40 ($-0.858$ bohr vs $-0.023$ bohr).

As we will see in the next section, however,  
the gyration tensor coefficients of the unstrained phase 
do not anticipate the strength of the 
piezogyration response in the crystal.
%

\subsection{Piezogyration}\label{Sec_elasto}
The numerical results for the calculated linearly independent components of both the piezogyration and elastogyration tensors
are shown in Table~\ref{Tab_piezo} and \ref{Tab_elasto}, respectively. 
We provide in Appendix~\ref{App_strategy} a detailed explanation of the 
computational strategy; Appendix~\ref{Sec_app_elastic} provides the elastic stiffness and compliance tensors.
\begin{table}[t!]
	\caption{Independent components of the piezoogyration tensor (in $10^{-2}$ bohr/GPa) for the considered ternary 
		ABC$_2$ chalcopyrite crystals.
	}
	\label{Tab_piezo}
	\begin{ruledtabular}  
		\begin{tabular}{lrrrrr}
			&$\Pi_{11}$&$\Pi_{21}$&$\Pi_{31}$&$\Pi_{13}$&$\Pi_{55}$\\\hline 
			AgGaS$_2$ &3.31&42.32&$-48.00$&37.63&$-0.06$\\
			CuAlSe$_2$&$-1.36$&45.75&$-44.60$&$45.30$&0.84\\[4pt]
			ZnSiP$_2$ &$-0.66$&19.13&$-18.29$&18.59&1.63\\
			ZnGeP$_2$ &$-0.71$&33.52&$-31.03$&32.26&1.50 \\
			ZnSnP$_2$  &0.74&57.19&$-59.47$&58.03&$-1.76$\\[4pt]
			MgSiP$_2$ &$-0.04$&5.78&$-5.37$&4.46&2.18 \\
			MgGeP$_2$ &1.64&18.00&$-14.45$&14.48&2.46 \\
			MgSnP$_2$ &3.69&34.38&$-33.31$&30.81&1.19 \\
		\end{tabular}
	\end{ruledtabular}
\end{table}

\begin{table}[t!]
	\caption{Independent components of the elastogyration tensor (in bohr) for the considered ternary 
		ABC$_2$ chalcopyrite crystals.
	}
	\label{Tab_elasto}
	\begin{ruledtabular}  
		\begin{tabular}{lrrrrr}
			&$R_{11}$&$R_{21}$&$R_{31}$&$R_{13}$&$R_{55}$ \\\hline 
			AgGaS$_2$&$-0.87$&14.03&$-13.85$&7.74&$-0.02$\\
			CuAlSe$_2$&$-1.15$&18.30&$-17.23$&15.51&0.35\\[4pt]
			ZnSiP$_2$ &$-0.50$&15.28&$-14.63$&12.94&1.18\\
			ZnGeP$_2$&$-1.30$&24.82&$-22.25$&21.49&1.00 \\
			ZnSnP$_2$ &1.16&29.75&$-31.73$&33.41&$-0.94$\\[4pt]
			MgSiP$_2$ &$-0.26$&4.24&$-3.74$&$-0.29$&1.10\\
			MgGeP$_2$ &0.93&9.91&$-7.98$&1.96&1.11\\
			MgSnP$_2$ &1.99&11.27&$-11.60$&9.14& 0.44\\
		\end{tabular}
	\end{ruledtabular}
\end{table}
Let us focus on the elastogyration coefficients reported in Table~\ref{Tab_elasto}; much of the discussion can be extended to
the piezogyration case as reported in Table~\ref{Tab_piezo} once elasticity is included via Eq. (\ref{Eq_T_from_R}).
We note that $\abs{R_{11}}$ and $\abs{R_{55}}$ are consistently smaller than 
$\abs{R_{21}}$, $\abs{R_{31}}$, and $\abs{R_{13}}$ by one to two orders of magnitude across all compounds.
Moreover, the numerical results reveal the approximate relation $R_{21}\sim -R_{31}\sim R_{13}$.
These observations highlight the close structural relationship between the ternary ABC$_2$ crystal structure
and the zinc-blende structure with space group $F\bar{4}3m$, as discussed in Sec.~\ref{Sec_strc}.
In particular, the $\bar{4}3m$ point-group symmetry imposes the exact relations 
$R_{11}=R_{55}=0$ and $R_{21}=-R_{31}=R_{13}$, thereby 
leaving only a single linearly-independent elastogyration coefficient
and ensuring $\Tr(g)=0$ at first-order in the applied strain.

Additionally, we note that $R_{21}>0$, $R_{31}<0$, and $R_{13}>0$ hold consistently 
in all studied crystals. Their relative signs are  
inherited from the symmetry relations of the ideal zinc-blende structure with $\bar{4}3m$ symmetry, as 
previously discussed. The absolute sign of each entry, however, deserves further exploration.
Let us focus exclusively on $R_{31}$. (From Eq.~(\ref{Eq_T_from_R}) and invoking the
crystal symmetries of the $\bar{4}2m$ point group, we deduce that
 $\Pi_{31}=R_{31}(\mathcal{S}_{11}-\mathcal{S}_{12})$, and
mechanical stability require $\mathcal{S}_{11}-\mathcal{S}_{12}=1/(\mathcal{C}_{11}-\mathcal{C}_{12})>0$, meaning that 
the sign
of $\Pi_{31}$ follows directly from that of $R_{31}$.)
What imposes $R_{31}<0$ in all the ternary chalcopyrite crystals we have considered? 
We hypothesize 
that the sign of $R_{31}$ is determined by the orientation of the cation-anion tetrahedra together
with the sense of the bond polarity, which is the same for all compounds studied here, while 
additional atomic details mainly affect its magnitude. 
(This is supported by the computational experiment we perform in Appendix~\ref{App_Si}, 
in which the cation-anion asymmetry is switched on continuously from Si $\to$ AlP $\to$ MgSiP$_2$ without any change 
of sign in $R_{31}$.)

\begin{figure}[t]
	\includegraphics[width=\columnwidth]{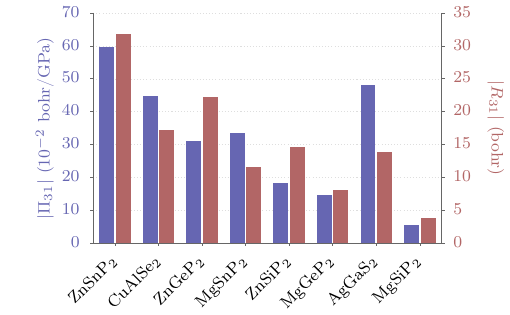}
	\caption{Comparison of the $\abs{\Pi_{31}}$ and $\abs{R_{31}}$ tensor entries among the studied 
		materials, ordered from left to right with increasing $\frac{1}{4}-u$.}
	\label{fig_crystals}
\end{figure}

Figure.~\ref{fig_crystals} compares the magnitude of $\abs{\Pi_{31}}$
and $\abs{R_{31}}$ across the studied crystals, ordered 
from left to right with increasing $\frac{1}{4}-u$, where the value of the internal parameter $u$ that controls 
the ionic displacement is shown in Table~\ref{Tab_strc} for each compound. According to our
analysis in Appendix~\ref{App_Si} on MgSiP$_2$  (the material exhibiting larger distortion from the zinc-blende parent), $u$ is the structural parameter responsible for most
of the renormalization of $R_{31}$, while the tetragonal distortion $\eta$
enters only as a minor correction.
Fig.~\ref{fig_crystals} reveals two main trends, formed by the
Zn$B$P$_2$ and Mg$B$P$_2$ series (with $B$=Si, Ge, Sn) of the II-IV-V$_2$
family, in  which $\abs{R_{31}}$ decreases monotonically with increasing
$\frac{1}{4}-u$. The two members of the I-III-VI$_2$ family, CuAlSe$_2$ and
AgGaS$_2$, seem to follow a similar behavior, although two data points are
not enough to draw general conclusions.

The ordering of $\abs{\Pi_{31}}$ differs
from that of $\abs{R_{31}}$ because
of the elastic compliance factor $\mathcal{S}_{11}-\mathcal{S}_{12}$, which 
is particularly large in AgGaS$_2$.

For completeness, we also examine how the uniaxial stress
$\sigma_1$---which induces $\Pi_{31}$---affects the 
internal atomic positions.
In the unstrained phase, a single internal parameter $u$ fixes all atomic positions (see Table ~\ref{Tab_wyck}).
Under $\sigma_1$, the symmetry is lowered to space group $I2_12_12_1$, which introduces additional 
structural degrees of freedom, as listed in Table~\ref{Tab_wyck_strain}. At linear order in the applied stress field, we define
the following coefficients,
\begin{equation}\label{Eq_Delta_deriv}
\begin{split}
\Delta\delta_X &=\frac{\partial \delta_X}{\partial \sigma_1}\bigg|_{\mathbf{\sigma=0}},\quad  \text{ for } X=\{A, B\},\\[6pt]
\Delta u_{xy}&=\frac{1}{2}\frac{\partial (u_x-u_y)}{\partial\sigma_1}\bigg|_{\boldsymbol{\sigma=0}}.
\end{split}
\end{equation}
which are also reported in Table~\ref{Tab_strc_stress}.
(We combine the individual displacements $u_x$ and $u_y$ into $(u_x-u_y)/2$,
so that the obtained quantity is odd in stress and strain; this will be useful in the context of the next section.)
As in the case of $R_{31}$,
we observe a consistent sign pattern across all materials for $\Delta\delta_A$, $\Delta\delta_B$ 
and $\Delta u_{xy}$. This further supports the view that, at least 
for the compounds studied here, the sign 
of $R_{31}$ is structural in origin, while electronic and chemical details modulate its magnitude without 
changing its sign. 

\begin{table}[t!]
	\caption{Atomic positions of the ternary chalcopyrite structure ABC$_2$
		under uniaxial stress along $x$ (space group $I2_12_12_1$), in the
		conventional body-centered orthorhombic cell containing 16 atoms, with
		lattice vectors $\mathbf{a}_1=a\,\hat{\mathbf{x}}$,
		$\mathbf{a}_2=b\,\hat{\mathbf{y}}$, $\mathbf{a}_3=c\,\hat{\mathbf{z}}$
		($a\neq b\neq c$, since the stress $\sigma_1$ is applied under zero stress condition, i.e., $\sigma_2=\sigma_3=0$). 
		Every site is to be taken together with the
		translation $(\frac{1}{2},\frac{1}{2},\frac{1}{2})$. The loss of the
		$\bar{4}$ symmetry axis splits the anion (C) orbit into two, with independent
		parameters $u_x$ and $u_y$, and releases the $z$ coordinates of the
		cations A and B, via the parameters $\delta_A$ and $\delta_B$. The unstrained structure of
		Table \ref{Tab_wyck} is recovered for $a=b$, $\delta_A=\delta_B=0$ and
		$u_x=u_y=u$.}
	\label{Tab_wyck_strain}
	\begin{ruledtabular}
		\begin{tabular}{ll}
			Atom &  Reduced coordinates\\ \hline \noalign{\vskip 2pt}
			A &  $(0,0,\delta_A)$,\quad $(0,\tfrac12,\tfrac14-\delta_A)$\\[4pt]
			B & $(0,0,\tfrac12+\delta_B)$,\quad $(0,\tfrac12,\tfrac34-\delta_B)$\\[4pt]
			C$_x$  & $(u_x,\tfrac14,\tfrac18)$,\quad $(1-u_x,\tfrac34,\tfrac18)$\\[4pt]
			C$_y$ & $(\tfrac14,\tfrac12+u_y,\tfrac38)$,\quad $(\tfrac34,\tfrac12-u_y,\tfrac38)$
		\end{tabular}
	\end{ruledtabular}
\end{table}

\begin{table}[t!]
	\caption{Derivatives of structural parameters $\delta_A$, $\delta_B$ and $\frac{1}{2}(u_x-u_y)$ with 
		respect to applied uniaxial stress along the Cartesian direction $x$ as defined
		 by Eq.~(\ref{Eq_Delta_deriv}). Values are given in $10^{-2}$ bohr$/$GPa.
	}
	\label{Tab_strc_stress}
	\begin{ruledtabular}  
		\begin{tabular}{lrrr}
			&$\Delta \delta_A$&$\Delta \delta_B$&$\Delta u_{xy}$ \\\hline 
			AgGaS$_2$ &5.343 &$-2.735$ & 7.455\\
			CuAlSe$_2$ &0.651 &$-0.790$ & 4.217\\[4pt]
			ZnSiP$_2$ &0.614 &$-0.411$ &2.192 \\
			ZnGeP$_2$ &0.618 & $-0.416$&2.495 \\
			ZnSnP$_2$ &0.478 &$-0.464$ & 3.253\\[4pt]
			MgSiP$_2$ & 1.141 & $-0.709$ & 2.976 \\
			MgGeP$_2$ & 1.489& $-0.924$& 3.924\\
			MgSnP$_2$ &2.414 &$-1.753$ &6.443 \\
		\end{tabular}
	\end{ruledtabular}
\end{table}

%


\subsection{Piezochirality}\label{Sec_elastochiral}
Mechanical control of structural handedness and of the associated optical activity 
has been reported for decades~\cite{doi:10.1143/JPSJ.43.195,Wadhawan:a17058,Vlokh01071994,Martynyuk-Lototska2010,ETCHEGOIN1992655},
and strain has more recently been used to drive and stabilize transitions 
into chiral phases~\cite{PhysRevB.111.174102,ren2025strain,PhysRevLett.107.117602}. 
A general formalism of the strain-chirality coupling, however, was only established recently in Ref.~\cite{Zeng2026}, 
where the authors show via group-theoretical symmetry analysis 
which crystal symmetries allow linear---and higher order---piezochiral coupling.
Crucially, the crystal that Ref.~\cite{Zeng2026} employs in order to test that strain induces chirality at linear order in the applied strain field is AgGaS$_2$, one of the 
chalcopyrite crystals considered here and the textbook material for non-enantiomorphic yet optically active crystals~\cite{Hobden1967,hobden1968optical}. Structural chirality is inferred in Ref.~\cite{Zeng2026}  by measuring 
the strain-induced optical rotation of light propagating along the crystal's principal $c$-axis, 
i.e., the optical rotation induced by $g_3$.

Whether the \textit{magnitude} of the induced gyration component $g_3$ is directly related to
structural chirality is, however, an open question, in particular because there is no
clear consensus on how to quantify chirality itself~\cite{Zabrodsky1995CSMChirality,RevModPhys.71.1745,doi:10.1139/v99-223,PhysRevB.110.174112,https://doi.org/10.1002/jcc.20990, spaldin2026}.
On top of that, we should bear in mind that 
optical activity is a response function and symmetry 
determines whether it can occur, but its magnitude and sign 
cannot be determined by structure alone~\cite{Glazer1986}. Natural optical activity 
can nevertheless help distinguish chiral from achiral crystals.
There is, in fact, one contribution from the gyration tensor that
encodes information about chirality, is linear in the tensor components and remains
invariant irrespective of the reference frame: its trace.
As previously discussed at the end of Sec.~\ref{sec_point_group}, $\Tr(g)\neq 0$ only if the system is chiral.
Motivated by this observation,
we define the following quantity in terms of either the piezogyration or elastogyration coefficients,
\begin{equation}\label{Eq_C_tr}
	\Upsilon_X=\frac{1}{3}\left( X_{11}+ X_{21} + X_{31} \right), 
	\quad X\in\{\Pi,\,R,\,R^\sigma\},
\end{equation}
which measures the induced changes in $\Tr(g)$ due to a uniaxial strain (or stress) along the Cartesian direction $x$.
$\Upsilon_\Pi$ and $\Upsilon_R$ are defined through the piezogyration and elastogyration coefficients
reported in Sec. \ref{Sec_elasto}. On the other hand, $\Upsilon_{R^\sigma}$ is written in
terms of the $R^\sigma_{IJ}$ coefficients defined 
in Appendix \ref{App_strategy}, which capture strain derivatives of the gyration tensor under zero stress 
condition (see Appendix~\ref{App_strategy} for more details).
We report in Table~\ref{Tab_piezochiral} the obtained numerical values for the three \textit{flavors} of $\Upsilon$
defined by Eq.~(\ref{Eq_C_tr}). 
Note that the relationship among these three quantities is dictated by elasticity through
\begin{equation}\label{Eq_upsilon}
\begin{split}
	\Upsilon_R &= (\mathcal{C}_{11}-\mathcal{C}_{12})\,\Upsilon_\Pi,\\
	\Upsilon_{R^\sigma} &= (1+\nu_{12})\,\Upsilon_R.
	\end{split}
\end{equation}
The first relation involves the single elastic stiffness combination
$\mathcal{C}_{11}-\mathcal{C}_{12}=1/(\mathcal{S}_{11}-\mathcal{S}_{12})$,
whereas the second
depends on the Poisson ratio $\nu_{12}$ alone, since both $R$ and
$R^\sigma$ are derivatives with respect to strain (see Appendix~\ref{App_strategy}).
\begin{table}[t!]
	\caption{Piezochiral effect as measured by the derivative of the trace of the gyration tensor with respect to the external field, either strain or stress. Three ``flavors'' of this quantity are presented, in terms of the
		 tensor quantities $\boldsymbol{\Pi}$, $\mathbf{R}$ and $\mathbf{R}^\sigma$, where $\Upsilon_\Pi$ is given in $10^{-2}$ bohr/GPa units
		 and both $\Upsilon_R$ and $\Upsilon_{R^\sigma}$ in bohr.}
	\label{Tab_piezochiral}
	\begin{ruledtabular}
		\begin{tabular}{lrrr}
			&$\Upsilon_\Pi$&$\Upsilon_R$&$\Upsilon_{R^\sigma}$\\\hline
			AgGaS$_2$&$-0.79$&$-0.23$&$-0.32$\\
			CuAlSe$_2$ &$-0.07$&$-0.03$&$-0.04$\\[4pt]
			ZnSiP$_2$ &0.06&0.05&0.06\\
			ZnGeP$_2$ &0.59&0.42&0.55\\
			ZnSnP$_2$ &$-0.52$&$-0.28$&$-0.37$\\[4pt]
			MgSiP$_2$&0.12&0.08&0.08\\
			MgGeP$_2$&1.73&0.95&1.08\\
			MgSnP$_2$&1.59&0.55&0.73
		\end{tabular}
	\end{ruledtabular}
\end{table}
From Table~\ref{Tab_piezochiral}, the first thing we note is that, unlike the elastogyration coefficient
$R_{31}$ and the structural parameters $\Delta\delta_A$, $\Delta\delta_B$ and $\Delta u_{xy}$ defined in the previous section, 
the sign of $\Upsilon$ depends on the material. This is something we could have foreseen: for the cases we have studied, 
$R_{31}\sim -R_{21}$ is determined by the large piezogyration inherited from the parent zinc blende structures,
and the sign of these coefficients is the same across materials.
The sign of $R_{11}$,
by contrast, depends on additional chemical details and, as Table~\ref{Tab_elasto} shows, can be either positive or negative. Similarly, the relative magnitude of $\abs{R_{31}}$ and $\abs{R_{21}}$ is 
material dependent.
Consequently, $\Upsilon$ is a small residual that arises from the near cancelation of large contributions, and its sign
varies even among crystals with similar structural properties.
We would like to stress that $\Upsilon$, which encodes information about strain-induced 
changes on $\Tr(g)$, is not a measure of
chirality. It represents one pseudoscalar among
(infinitely) many~\cite{RevModPhys.71.1745} and can vanish accidentally
even in chiral systems~\cite{pelloni2017electric}. 
$\Tr(g)\neq 0$ is therefore a 
sufficient but not necessary condition for chirality, and provides no information about
the \textit{amplitude} of the chiral distortion.

\begin{figure}[t]
	\includegraphics[width=\columnwidth]{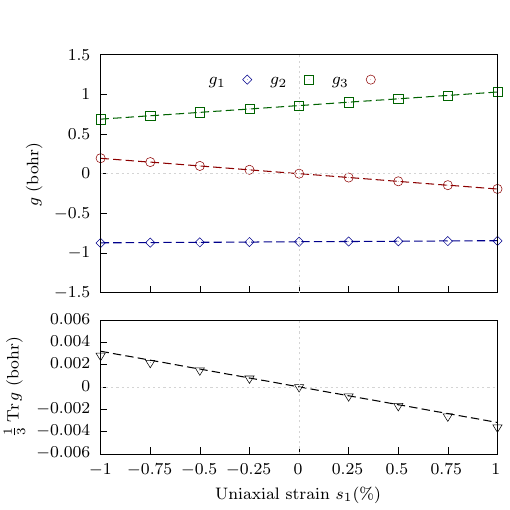}
	\caption{The top panel shows the uniaxial-strain dependence of the gyration tensor components $g_1$, $g_2$ and $g_3$ 
		in AgGaS$_2$. The dashed lines 
		indicate $g_{I}=g_I^{(0)}+R^\sigma_{I1}s_{1}$, where $g_I^{(0)}$ are the gyration tensor components
		in the unstrained phase, and the $R^\sigma_{IJ}$ coefficients are reported in Table \ref{Tab_tilde_elasto}.
		The lower
		panel shows the isotropic average $\frac{1}{3}\,\Tr(g)$, where the dashed line is obtained from
		the $\Upsilon_{R^\sigma}$ coefficient reported in Table~\ref{Tab_piezochiral}.}
	\label{fig_AgGaS2}
\end{figure}

\begin{figure}[t]
	\includegraphics[width=\columnwidth]{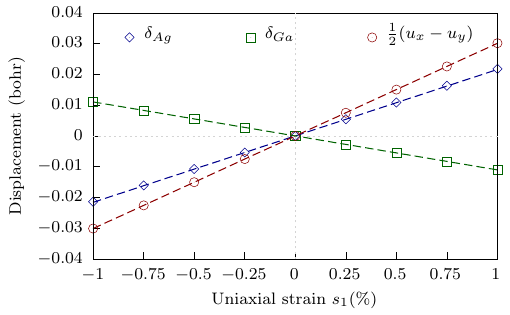}
	\caption{Uniaxial-strain dependence of the internal structural parameters in AgGaS$_2$, 
		defined in Table~\ref{Tab_wyck_strain}. The dashed lines are obtained from the linear coefficients given by
		 Eq. (\ref{Eq_Delta_deriv}). Here, however, they are expressed as strain derivatives, using the scaling factor $(1+\nu_{12})(\mathcal{C}_{11}-\mathcal{C}_{12})$, consistent with Eq. (\ref{Eq_upsilon}).}
	\label{fig_derivs}
\end{figure}

As a representative test case of the systems we have considered, and in light of the 
recent study on  piezochirality in silver gallium sulfide~\cite{Zeng2026}, we provide 
in Fig.~\ref{fig_AgGaS2} and 
Fig.~\ref{fig_derivs}, respectively,  the evolution with applied strain
of the independent components of the gyration tensor of AgGaS$_2$, together with its trace, and of the
internal structural degrees of freedom as defined in Table~\ref{Tab_wyck_strain}. The data points of these figures are computed
for a uniaxial strain $s_I\equiv s_{xx}$ under zero stress conditions, i.e.,
$\sigma_J=0$, $\forall J\neq I$ (see Appendix~\ref{App_strategy}), because this is the
 situation one would encounter in experiments. The dashed lines are obtained from the linear
coefficients we have already discussed
(more details are given in the captions of
 Fig.~\ref{fig_AgGaS2} and Fig.~\ref{fig_derivs}).

Our calculated linear coefficients properly describe the strain induced 
changes in the gyration tensor and internal structural parameters within
the considered uniaxial strains between [$-1\%,+1\%$], as shown in Fig.~\ref{fig_AgGaS2} 
and Fig.~\ref{fig_derivs}. 
We note that the experimental strain amplitude considered in Ref.~\cite{Zeng2026} is 50 times smaller. 
For an applied uniaxial strain of $0.0207\%$, Ref.~\cite{Zeng2026} reports an optical rotatory power 
of $\rho\sim -0.429$ deg/mm for light traveling parallel to the $c$-axis, 
probed with light with a wavelength of $\lambda=808$ nm (corresponding to 
$\hbar\omega\sim 1.53$ eV, well within the transparent regime). This gives 
$\rho(\omega)/(\hbar\omega)^2\sim -0.18$ deg/[mm(eV)$^2$],
which is about $4\%$ of 
the zero-frequency limit of the rotatory power of 
$\alpha$-SiO$_2$, with a measured value of $\bar{\rho}\sim 4.6$ deg/[mm(eV)$^2$]~\cite{PhysRevLett.76.1372}.
From our calculations, the rotatory power for light traveling parallel to the $c$-axis induced
by a uniaxial strain along $x$ is obtained through the $R_{31}^\sigma$ coefficients from 
Table~\ref{Tab_tilde_elasto}. For a  strain amplitude of $s_1=0.0207\%$, 
we obtain $\bar{\rho}= -0.16$ deg/[mm(eV)$^2$], 
in excellent agreement with the experiment. 
Before closing this section,
we would like to caution against using a single optical rotation 
measurement to establish whether a structure is chiral.
To this end, we split the symmetric gyration tensor $g_{(ij)}$ defined in 
Eq.~(\ref{Eq_sym_antisym}) 
into an isotropic 
and a symmetric traceless part,
\begin{equation}
 g_{(ij)} = \frac{1}{3}\,\mathrm{Tr}(g)\,\delta_{ij}
              +\tilde{g}_{(ij)},
\label{eq_decomp}
\end{equation}
where $\Tr(\tilde{g})=0$ by definition, and $\Tr(g)\neq 0$ only if the system is chiral.
The trace is the only linear contribution of optical activity that survives
orientational averaging and, 
as a consequence, optical rotation is measurable for molecules in solution 
only if the molecules themselves are chiral.
The traceless part $\tilde{g}_{(ij)}$, however, also gives rise to optical rotation and is allowed in achiral
gyrotropic point groups, as shown by the unstrained chalcopyrites themselves where
gyration tensor in Eq.~(\ref{Eq_g_vec}) is traceless.
In the situation considered above, namely AgGaS$_2$ strained along $x$,
a nonzero $g_{33}$ unequivocally establishes that the crystal is chiral
only if $g_{11}$ and $g_{22}$ are also known to be nonzero,           
so that the trace is known.
The parent zinc-blende structure provides an illustrative example.
In the $\bar{4}3m$ point group, the natural optical activity vanishes identically, i.e.,
$g_{ij}=0,\,\forall i,j$.
A uniaxial strain $s_{1}$ lowers the symmetry to $\bar{4}2m$, with the $\bar{4}$ axis now
along $x$. This point group is achiral but gyrotropic, and the zinc-blende relations
$R_{11}=0$ and $R_{21}=-R_{31}$ discussed in Sec.~\ref{Sec_elasto} give
\begin{equation}\label{Eq_g_zb_strain}
	\Delta g=R_{31}\,s_1
	\begin{pmatrix}
		0&0&0\\
		0&-1&0\\
		0&0&1
	\end{pmatrix}.
\end{equation}
A uniaxial strain along $x$ therefore induces, at linear order, optical rotation for light
propagating along $z$ in a zinc-blende crystal (see also Ref.~\cite{ETCHEGOIN1992655}),
while the crystal remains achiral and $\Tr(g)=0$.

\section{Conclusions and outlook}\label{sec_conclusions}
We have presented a first-principles study of the piezogyration and elastogyration
effects in eight representative ternary chalcopyrite crystals with ABC$_2$ stoichiometry
belonging to the I-III-VI$_2$ and II-IV-V$_2$ families. The computational methodology for 
the piezogyration (elastogyration) presented here 
is general and combines the DFPT implementation for the natural optical activity publicly available 
in the 
\textsc{abinit} code~\cite{Verstraete2025,Zabalo2023} with central finite differences on applied strain (stress) fields.
Several independent routes to the same coefficients, together
with the agreement between the elastic compliances and stiffness components 
extracted from the relaxed cells and those computed directly with DFPT, 
confirm the internal consistency of the approach 
(see Appendix~\ref{App_strategy} and \ref{Sec_app_elastic}).

Our results reveal that the elastogyration of the ternary chalcopyrites is largely a property 
inherited from their parent zinc-blende structures, with $\bar{4}3m$ point group symmetry. This point group 
allows for a single independent elastogyration coefficient, which is at the origin of the 
approximate relations $R_{11}\sim R_{55}\sim 0$ and $R_{21}\sim -R_{31}\sim R_{13}$ that 
the chalcopyrite's elastogyration coefficients fulfill. Furthermore, the sign of $R_{31}$ is the same
in all the cases we have studied.  We explain this fact by noting that
the orientation of the cation-anion tetrahedra, together with the sense of the bond polarity,
is common to all the chalcopyrites studied here. This claim is supported by the 
computational experiment we perform in Appendix~\ref{App_Si}, where a  
Si crystal with diamond-like structure is taken as the starting point 
and continuously transformed along the path
Si $\to$ AlP $\to$ MgSiP$_2$ through alchemical 
mixing, while $R_{31}$ preserves its sign throughout.
We note that the optical activity 
of the unstrained phase does not constitute a good proxy for predicting the magnitude 
of the elastogyration: ZnSnP$_2$ exhibits 
only modest natural optical activity,
however, it displays the largest $\abs{R_{31}}$
among the compounds considered.
The piezogyration coefficients, on the other hand, 
depend on additional elastic properties, which can decrease or increase the response;  
AgGaS$_2$ is a clear example of the latter.

As an outlook, it will be interesting to extend the methodology presented here
to spatially inhomogeneous stress and strain field effects on the gyration tensor, 
i.e., gradient piezogyration and gradient elastogyration, defined
in terms of nonuniform stress and strain fields, 
respectively~\cite{Vasylkiv2013,skab2011optical,vasylkiv2013optical,Vlokh1997}. 
Unlike the piezogyration (elastogyration), the gradient piezogyration (elastogyration) tensor can be nonzero even in centrosymmetric systems. 
While the pioneering studies of Vlokh and co-workers established much of the experimental groundwork in this context~\cite{Vasylkiv2013,skab2011optical,vasylkiv2013optical,Vlokh1997}, 
first-principles calculations are still missing.

Another promising, and so far unexplored, research direction
is gradient piezochirality, or \textit{flexochirality}, namely
chirality induced by spatially inhomogeneous strain (or stress) fields.
By analogy with homogeneous piezogyration, this effect could be 
captured through the effects of strain gradients on the trace of the gyration tensor.
%
The situation is, in many aspects, similar to that of piezochirality,
whose symmetry requirements had been known for years before the effect was 
formalized only recently~\cite{Zeng2026}. We expect that the field of \textit{flexochirality}      
will follow a similar path, and we hope our work stimulates further theoretical 
research about strain-induced effects, both uniform and nonuniform, on natural optical activity.

\begin{acknowledgments}
E.B. and A.Z. acknowledge the Fonds de la Recherche Scientifique (FNRS) PDR project CHRYSALID No.40003544, the Excellence of Science program (EOS ``ShapeME'', No. 40007525) funded by the FWO and F.R.S.-FNRS, sthe Consortium des \'Equipements de Calcul Intensif (C\'ECI), 
funded by the F.R.S.-FNRS under Grant No. 2.5020.11 and the Tier-1 Lucia supercomputer of the Walloon Region, infrastructure funded by the Walloon Region under the grant agreement No. 1910247. A.H.R. acknowledges the support from NSF-OAC-2513657, a project funded by the National Science Foundation.
F.G.O. acknowledges financial support from the Fonds de la Recherche Scientifique (FNRS) through the grant FNRS-CR 1.B.227.25F

\end{acknowledgments}

\begin{appendices}

\section{Computational strategy}\label{App_strategy}

\begin{table}[t!]
	\caption{Derivatives of the gyration vector with respect to the applied strain under uniaxial-stress conditions, 
		$R^\sigma_{IJ}$, as described by Eq.~(\ref{Eq_tilde_R}), given in bohr units. This is the configuration realized when a free-standing crystal is loaded along a single direction and the strain is measured directly, and it is neither the elastogyration tensor  defined at fixed remaining strain, nor the piezogyration tensor; the three are related by 
		$R^\sigma_{IJ} = \Pi_{IJ}/S_{JJ}$ (no summation on $J$) and $R_{IJ} = \Pi_{IK} C_{KJ}$. Unlike $R_{ijkl}\equiv R_{IJ}$ and $\Pi_{ijkl}\equiv \Pi_{IJ}$, with $ij\to I$ and 
        $kl\to J$, $R^\sigma_{IJ}$ is not a fourth-rank tensor and its components refer specifically to the crystallographic axes of Table~\ref{Tab_strc}. 
        It has nonetheless the same matrix form as $R_{IJ}$, owing to the equal
diagonal compliances of the $\bar{4}2m$ class, namely  $\mathcal{S}_{11}=\mathcal{S}_{22}$ and
$\mathcal{S}_{44}=\mathcal{S}_{55}$; this need not hold in trigonal or hexagonal classes.
	}
	\label{Tab_tilde_elasto}
	\begin{ruledtabular}  
		\begin{tabular}{rrrrrr}
			&$R^\sigma_{11}$&$R^\sigma_{21}$&$R^\sigma_{31}$&$R^\sigma_{13}$&$R^\sigma_{55}$ \\\hline 
			AgGaS$_2$& 1.34&17.09&$-19.39$&13.60&$-0.02$\\
			CuAlSe$_2$&$-0.71$&24.02&$-23.42$&23.09&0.35\\[4pt]
			ZnSiP$_2$ &$-0.68$&19.60&$-18.74$&18.12&1.18\\
			ZnGeP$_2$ &$-0.66$&31.31&$-28.99$&29.76&1.00\\
			ZnSnP$_2$ &0.53&40.97&$-42.61$&42.99&$-0.94$\\[4pt]
			MgSiP$_2$ &$-0.03$&4.00&$-3.72$&1.36&1.10\\
			MgGeP$_2$ &1.03&11.27&$-9.05$&5.25&1.11\\
			MgSnP$_2$ &1.70&15.82&$-15.33$&12.69&0.44\\
		\end{tabular}
	\end{ruledtabular}
\end{table}
For computational convenience, we take strain, rather than stress, as the externally controlled field.
In particular, we apply $\pm 0.5\%$ strain amplitudes in all of our calculations, and compute the derivatives with respect 
to strain (or stress)
via central finite differences.
Note that the definition for the elastogyration coefficient given by Eq.~(\ref{Eq_elasto})
fixes all strain components other than the one being varied. In practice, however, a crystal subjected to
a uniaxial strain along $x$, for example, is free to contract or expand in the transverse directions $y$ and $z$, 
such that $\sigma_1$ is the only 
nonzero stress component. This 
is also the configuration that we find most convenient for our first-principles calculations, since it only requires
that the residual transverse stresses be relaxed away. Such computational setup enables extracting directly the piezogyration coefficients
as defined by Eq.~(\ref{Eq_piezo_def}); the elastogyration tensor is then obtained through the relations given
by Eq.~(\ref{Eq_T_from_R}). 

Linear elasticity of the crystals is also readily accessible through this 
strategy. Consider, for example, a uniaxial strain $s_1$ applied along the Cartesian direction $x$, with $s_2$ and $s_3$ allowed
to relax until $\sigma_2=\sigma_3=0$. Along this path, and within the linear elastic regime,
the transverse strains $s_2$ and $s_3$ are subject to the Poisson 
ratios of the unstrained crystal,
\begin{equation}
	\nu_{12}=-\frac{d s_2}{d s_1}\bigg|_{\sigma_2=\sigma_3=0},\quad 
	\nu_{13}=-\frac{d s_3}{d s_1}\bigg|_{\sigma_2=\sigma_3=0}.
\end{equation}
Similarly, for a uniaxial strain $s_3$ applied along the Cartesian direction $z$, with $s_2$ and $s_3$ allowed to evolve 
until $\sigma_1=\sigma_2=0$, we define the following  Poisson ratio
\begin{equation}
	\nu_{33}=-\frac{d s_1}{d s_3}\bigg|_{\sigma_1=\sigma_2=0} = -\frac{d s_2}{d s_3}\bigg|_{\sigma_1=\sigma_2=0}.
\end{equation} 
In fact, the (relaxed-ion) elastic compliance coefficients are accessible following a similar logic; we report
both the computed quantities obtained through the dilation and contraction of the lattice vectors under the discussed 
zero-stress condition, as well as the obtained values via density-functional perturbation theory (DFPT) 
in Sec.~\ref{Sec_app_elastic}, where the agreement between the two approaches is nearly perfect.

As a further consistency test, we determine the elastogyration coefficients using an alternative approach by introducing the following intermediate quantity,
\begin{equation}\label{Eq_tilde_R}
	R^\sigma_{IJ}=\frac{d \Delta g_I}{d s_J}\Big|_{\sigma_K=0,\forall K\neq J},
\end{equation}
which differs from the partial derivative definition of the elastogyration coefficients given by Eq.~(\ref{Eq_elasto})
precisely by the fact that the transverse components of strain are allowed to follow. We stress that 
$\mathbf{R}^\sigma$ is neither the elastogyration nor the piezogyration tensor: it is a derivative with respect to strain,
like the elastogyration $\mathbf{R}$, but taken under the boundary conditions that
define $\boldsymbol{\Pi}$. For completeness, 
we report in Table~\ref{Tab_tilde_elasto} the obtained numerical results
following Eq.~(\ref{Eq_tilde_R}).
The standard elastogyration tensor coefficients, as reported in Table~\ref{Tab_elasto} in the main text, are obtained through 
the following relationships, which are valid for the crystals with $\bar{4}2m$ point group symmetry,
\begin{equation}\label{Eq_Rtilde_R}
\begin{split}
R_{11}&=(\beta-\alpha)/2,\\
R_{21}&=(\beta+\alpha)/2,\\	
R_{31}&=R^\sigma_{31}/(1+\nu_{12}),\\
R_{13}&=R^\sigma_{13}-\nu_{33}\alpha,\\
R_{55}&=R^\sigma_{55},
\end{split}
\end{equation}
where we have defined 
\begin{equation}
\begin{split}
\alpha&=\frac{R^\sigma_{21}-R^\sigma_{11}-2\nu_{33}R^\sigma_{13}}{1-\nu_{12}-2\nu_{13}\nu_{33}},\\
\beta&=\frac{R^\sigma_{11}+R^\sigma_{21}}{1+\nu_{12}}.	
\end{split}
\end{equation}
Note that $R_{55}=R^\sigma_{55}$ in Eq.~(\ref{Eq_Rtilde_R}), indicating that a shear strain $s_5$ generates 
no stresses in addition to $\sigma_5$, and no transverse relaxation takes place at linear order. 
In addition, individual elastic stiffness or compliance coefficients do not enter into
the relationships of Eq.~(\ref{Eq_Rtilde_R}), since everything is expressed exclusively in terms 
of the Poisson \textit{ratios}. %
The zinc-blende relations discussed in Sec.~\ref{Sec_elasto} are
preserved by the transformation given by Eq.~(\ref{Eq_Rtilde_R}), since in the cubic limit $\nu_{12}=\nu_{13}=\nu_{33}$.

We finally note that the piezogyration tensor $\boldsymbol{\Pi}$ is recovered directly from $\mathbf{R}^\sigma$ via
$\Pi_{IJ}=R^\sigma_{IJ} S_{JJ}$, with no summation on the index $J$.

Overall, the computational strategy described here enables the calculation of the physically meaningful
piezogyration and elastogyration coefficients through several complementary routes:
 (i) piezogyration directly as the stress derivative of the gyration tensor, (ii) piezogyration from
  the $R^\sigma_{IJ}$ coefficients, (iii) elastogyration from piezogyration through 
Eq.~(\ref{Eq_T_from_R}), and (iv) elastogyration from the $R^\sigma_{IJ}$ coefficients. The agreement
between these routes further demonstrates the internal consistency and robustness of our implementation.

\section{On the origin of $R_{31}$}\label{App_Si}
This section is devoted to explain why the sign of the elastogyration coefficient $R_{31}$
is consistent across all the materials studied in this work. The hypothesis is that the
$R_{31}$ coefficient of the chalcopyrites is a property largely 
inherited from their parent zinc-blende structure,
where the structural distortions of the chalcopyrites modify its magnitude but not its sign.

We test this hypothesis by computing the elastogyration coefficient $R_{31}$ 
for a series of alloys along a continuous path from the diamond-structure 
Si crystal to the MgSiP$_2$ chalcopyrite,
which is the compound exhibiting not only
the largest deviation of the internal parameter $u$ from the ideal $1/4$ value, but also 
the largest tetragonal distortion (see Table~\ref{Tab_strc}). These two structural parameters
($\eta$ and $u$) are, in fact,  largely correlated. 

For simulating the alloys we 
employ the alchemical mixing approximation
as implemented in \textsc{abinit}~\cite{SCHAROCH2014358}.
We employ PBEsol pseudopotentials from pseudodojo (v0.4.1), but regenerated
without exchange-correlation (XC) nonlinear core corrections. In addition, 
we employ a pseudopotential with only two valence electrons for Mg, instead of the 
ten-electron pseudopotential used elsewhere in this work. This ensures that the number of occupied bands 
remains constant through the entire alchemical path from AlP to MgSiP$_2$ (see below).

\begin{figure}[t]
	\includegraphics[width=\columnwidth]{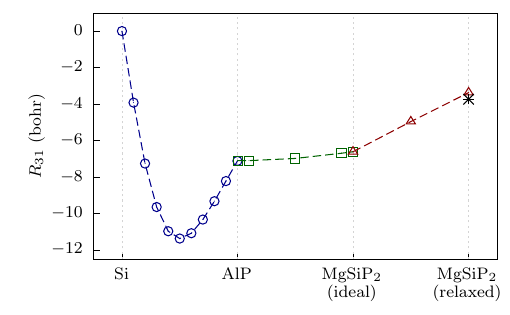}
	\caption{Elastogyration coefficient $R_{31}$ (clamped ions) along the
alchemical path Si $\to$ AlP $\to$ MgSiP$_2$.
We employ the alchemical mixing approximation
as implemented in \textsc{abinit} for simulating the alloys.
The first two segments, i.e., 
(Si$_{1-\lambda}$Al$_\lambda$)(Si$_{1-\lambda}$P$_{\lambda}$)
and
(Al$_{1-\lambda}$Mg$_\lambda$)(Al$_{1-\lambda}$Si$_\lambda$)P$_2$
with $\lambda=0 \to 1$, are
carried out at the ideal ($\eta=1$ and $u=\frac{1}{4}$) diamond-like geometry of Si,
for which a relaxed lattice parameter of $a=10.260$ bohr
is obtained for the conventional cell with 16 atoms as described by Table \ref{Tab_wyck}.
The third segment represents a linear interpolation of the parameters $a$, $c$ and $u$
from the ideal geometry of diamond Si to that of relaxed MgSiP$_2$
(Table~\ref{Tab_strc}).
The star is the $R_{31}$ elastogyration coefficient reported
in Table \ref{Tab_elasto} (with relaxed ions),
and helps visualize the effect of ionic relaxation.}
	\label{fig_alchemical}
\end{figure}

We start by considering the ideal diamond structure of the Si crystal, which 
is described in the setting of Table~\ref{Tab_wyck} with 
$\eta=1$, $u=\frac{1}{4}$ and A=B=C$\equiv$ Si.
We obtain a relaxed cell parameter
of $a=b=c/2=10.260$ bohr. The space group symmetry of diamond Si ($Fd\bar{3}m$) is centrosymmetric
and therefore
forbids the elastogyration effect, meaning that $R_{31}$ is trivially zero. 
While keeping the structure fixed, we then
consider the alchemical mixing Si$_{1-\lambda}$Al$_\lambda$ on the cation and
Si$_{1-\lambda}$P$_\lambda$ on the anion sublattice, reaching AlP at $\lambda=1$, where $\lambda$ controls the mixing.
This switches on the cation-anion asymmetry of the zinc-blende parent, 
and $R_{31}$ becomes nonzero. We find that $\abs{R_{31}}$ acquires its maximum
at $\sim \lambda=0.5$, as shown in the first segment of Fig.~\ref{fig_alchemical}.
Next, while still keeping the structural parameters to those of diamond Si, we 
follow the alchemical path (Al$_{1-\lambda}$Mg$_\lambda$)(Al$_{1-\lambda}$Si$_\lambda$)P$_2$.
This is shown in the
second segment of Fig.~\ref{fig_alchemical}, where little variation on
$R_{31}$ is observed. Finally, we perform a linear interpolation of the structural parameters 
$a,c$ and $u$, from the values of diamond Si (with $c=2a$ and $u=1/4$) to those of relaxed 
MgSiP$_2$ reported in Table~\ref{Tab_strc}. This step reduces the
$R_{31}$  elastogyration coefficient
of MgSiP$_2$
by about a half, from $-6.62$ bohr (ideal) to $-3.37$ bohr (relaxed), as 
shown in the third segment of Fig.~\ref{fig_alchemical}. 

To disentangle the individual effects of the $\eta$ and $u$ parameters, we perform two additional calculations.
Relaxing the lattice parameters $a,c$ while fixing $u=1/4$ we obtain $R_{31}=-6.57$ bohr, essentialy 
matching the ideal value. On the other hand, relaxing only the internal parameter $u$
while keeping the ideal cell ($c=2a$) yields $R_{31}=-4.00$ bohr. The reduction in magnitude of $R_{31}$ is therefore  mainly driven by $u$, which accounts for about $80\%$ of the renormalization, while the 
tetragonal distortion $\eta$ plays only a minor role.

The value we obtain at the end of the alchemical substitution, $R_{31}=-3.37$ bohr, 
is the clamped-ion elastogyration. This agrees with $10\%$ the relaxed-ion 
coefficient reported in Table~\ref{Tab_elasto}, which is indicated with a ``star'' symbol 
in Fig~\ref{fig_alchemical}. (Apart from the ionic relaxation effects,
we should not discard minor differences that may arise from the distinct pseudopotentials
used in either case: with vs without XC nonlinear core corrections, and ten vs two valence electrons for the Mg pseudopotential).

The conclusion of this section is that the sign of the elastogyration coefficient $R_{31}$ is
fixed at the first stage of the alchemical path: the additional structural distortions of the chalcopyrites, where $u$ appear to play the main driving force,
modify its magnitude but not its sign. We expect a similar behavior for analogous 
alchemical paths for the rest of the chalcopyrite crystals studied here.

\section{Elastic stiffness and compliance tensors}\label{Sec_app_elastic}
In this section we report both the elastic stiffness and compliance tensors of the crystals, 
calculated from the dilation and contraction of the lattice vectors under the applied strain fields
under zero stress condition. 
Calculations based on density-functional perturbation theory (DFPT), which are shown in parenthesis 
in Tables~\ref{Tab_compliance} and \ref{Tab_elastic},
are reported with a plane-wave cutoff energy of 50 Ha.
We find that the effects of the basis set truncation controlled by the plane-wave 
energy cutoff are 
more noticeable in the elastic properties than in the optical ones.
This is why we increased the cutoff energy from 40 Ha (used 
for all calculations in the main text) to 50 Ha here for the elastic properties 
computed via DFPT. We note that a similar convergence issue is also
present in the calculation of the stresses when the system is subject to a mechanical deformation.
However, both the piezogyration coefficients and the elastic compliances 
of Table~\ref{Tab_compliance} are computed from central finite differences of 
the stress, and these \textit{differences} are already converged at 40 Ha.

\begin{table*}[t!]
	\caption{Independent components of the relaxed-ion elastic compliance $\boldsymbol{\mathcal{S}}$ tensor 
			(in $10^{-2}$ GPa$^{-1}$ units), obtained from the dilation and contraction of the lattice vectors---zero stress condition---under the applied strain, where available (note that the $\mathcal{S}_{66}$ entry would require a $s_{xy}$ strain deformation that we did not perform, as it not needed for 
			building the whole piezogyration and elastogyration tensors). Values in brackets are obtained from a direct linear-response 
			calculation via DFPT.
	}
	\label{Tab_compliance}
	\begin{ruledtabular}  
		\begin{tabular}{lcccccc}
			&$\mathcal{S}_{11}$&$\mathcal{S}_{12}$&$\mathcal{S}_{13}$&$\mathcal{S}_{33}$&$\mathcal{S}_{44}$
			& $\mathcal{S}_{66}$\\\hline 
			AgGaS$_2$ &2.476 (2.477)&$-0.989$ ($-0.991$)&$-1.089$ ($-1.089$)&$2.766$ ($2.766$)&3.713 (3.730)&$\cdots$ (2.805)\\
			CuAlSe$_2$ &1.905 (1.910)&$-0.684$ ($-0.682$)&$-0.753$ ($-0.763$)&1.962 (1.955)&2.438 (2.441)&$\cdots$ (2.482)\\[4pt]
			ZnSiP$_2$ &0.976 (0.976)&$-0.274$ ($-0.274$)&$-0.337$ ($-0.337$)&1.026 (1.026)&1.387 (1.387)&$\cdots$ (1.419)\\
			ZnGeP$_2$ &1.070 (1.070)&$-0.325$ ($-0.324$)&$-0.343$ ($-0.343$)&1.084 (1.084)&1.512 (1.512)&$\cdots$ (1.525)\\
			ZnSnP$_2$ &1.396 (1.395)&$-0.479$ ($-0.478$)&$-0.453$ ($-0.452$)&1.350 (1.349)&1.870 (1.869)&$\cdots$ (1.872)\\[4pt]
			MgSiP$_2$ &1.444 (1.445) &0.007 (0.002)&$-1.207$ ($-1.202$)&3.283 (3.278)&1.990 (2.006)&$\cdots$ (2.815)\\
			MgGeP$_2$ &1.596 (1.600)&$-0.215$ ($-0.222$)&$-1.010$ ($-1.008$)&2.757 (2.751)&2.223 (2.242)&$\cdots$ (2.854)\\
			MgSnP$_2$ &2.173 (2.170)&$-0.697$ ($-0.696$)&$-0.930$ ($-0.928$)&2.427 (2.423)&2.712 (2.732)& $\cdots$ (3.038)\\
		\end{tabular}
	\end{ruledtabular}
\end{table*}	

\begin{table*}[t!]
	\caption{Independent components of the relaxed-ion elastic stiffness $\boldsymbol{\mathcal{C}}$ tensor (in GPa units), obtained from the dilation and contraction of the lattice vectors---zero stress condition---under the applied strain, where available. Values in brackets are obtained from a direct linear-response calculation via DFPT.
	}
	\label{Tab_elastic}
	\begin{ruledtabular}  
		\begin{tabular}{lcccccc}
			&$\mathcal{C}_{11}$&$\mathcal{C}_{12}$&$\mathcal{C}_{13}$&$\mathcal{C}_{33}$&$\mathcal{C}_{44}$
			& $\mathcal{C}_{66}$\\\hline 
			AgGaS$_2$ &94.0 (93.8)&65.1 (65.0)&62.7 (62.5)&85.5 (85.3)&26.9 (26.8)&$\cdots$ (35.6)\\
			CuAlSe$_2$ &97.2 (98.4)&58.5 (59.8)&59.8 (61.7)&96.9 (99.4)&41.0 (41.0)&$\cdots$ (40.3)\\[4pt]
			ZnSiP$_2$ &143.9 (144.0)&63.9 (64.0)&68.2 (68.2)&142.2 (142.3)&72.1 (72.1)&$\cdots$ (70.5)\\
			ZnGeP$_2$ &130.4 (130.5)&58.7 (58.7)&59.9 (59.9)&130.1 (130.2)&66.1 (66.1)&$\cdots$ (65.6)\\
			ZnSnP$_2$ &108.1 (108.2)&54.8 (54.8)&54.6 (54.6)&110.7 (110.8)&53.5 (53.5)&$\cdots$ (53.4)\\[4pt]
			MgSiP$_2$ &123.5 (123.1)&53.9 (53.8)&65.2 (64.8)&78.4 (78.0)&50.3 (49.9)&$\cdots$ (35.5)\\
			MgGeP$_2$ &105.6 (105.5)&50.4 (50.7)&57.2 (57.2)&78.2 (78.3)&45.0 (44.6)&$\cdots$ (35.0)\\
			MgSnP$_2$ &82.9 (82.9)&48.1 (48.0)&50.2 (50.1)&79.6 (79.7)&36.9 (36.6)& $\cdots$ (32.9)\\
		\end{tabular}
	\end{ruledtabular}
\end{table*}	

\begin{table}[t!]
	\caption{Poisson ratios (dimensionless) obtained from the relaxed-ion elastic stiffness tensor of Table~\ref{Tab_elastic}.
		These are defined as $\nu_{12}=-\mathcal{S}_{12}/\mathcal{S}_{11}$, 
		$\nu_{13}=-\mathcal{S}_{13}/\mathcal{S}_{11}$ and $\nu_{33}=-\mathcal{S}_{13}/\mathcal{S}_{33}$.
	}
	\label{Tab_Poisson}
	\begin{ruledtabular}  
		\begin{tabular}{lccc}
			&$\nu_{12}$&$\nu_{13}$&$\nu_{33}$\\\hline 
			AgGaS$_2$ &0.400 (0.400)& 0.440 (0.439)& 0.394 (0.394)\\
			CuAlSe$_2$ &0.359 (0.357)&0.396 (0.400)&0.390 (0.390)\\[4pt]
			ZnSiP$_2$ &0.281 (0.281)&0.345 (0.345)&0.328 (0.328)\\
			ZnGeP$_2$ &0.303 (0.303)&0.321 (0.320)&0.317 (0.316)\\
			ZnSnP$_2$ &0.343 (0.343)&0.324 (0.324)&0.335 (0.335)\\[4pt]
			MgSiP$_2$ &$-0.005$ ($-0.001$)&0.836 (0.832)&0.367 (0.367)\\
			MgGeP$_2$ &0.135 (0.139)&0.633 (0.630)&0.366 (0.366)\\
			MgSnP$_2$ &0.321 (0.321)&0.428 (0.427)&0.383 (0.383)\\
		\end{tabular}
	\end{ruledtabular}
\end{table}

\section{Experimental structural parameters}\label{app_exp}
The experimental structural parameters of the materials considered are collected in 
Table~\ref{Tab_strc_exp}, where available. These are to be compared with our calculated values of Table~\ref{Tab_wyck}. We note that MgSnP$_2$ has not been reported in experiments
to date. MgGeP$_2$ has been reported only in a 
cation-disordered zinc-blende phase with $a=5.652$\,\AA\,
\cite{FolberthPfister1961}.
\begin{table}[t!]
	\caption{Experimental lattice parameters $a=b$ and $c$ (in bohr),
    tetragonal distortion $\eta=c/2a$ and internal parameter
    $u$ (with respect to our setting of Table~\ref{Tab_wyck}), where available.}
	\label{Tab_strc_exp}
	\begin{ruledtabular}
		\begin{tabular}{lrrrrl}
			&$a$&$c$&$\eta$&$u$&Ref.\\\hline
			AgGaS$_2$  &10.880 & 19.471& 0.895& 0.291&  \cite{10.1063/1.1680242}\\
			CuAlSe$_2$ &10.601 &20.598 &0.971 &$\cdots$ & \cite{KUMAR20004}\\
			ZnSiP$_2$  &10.203 &19.719 &0.966 &0.269 & \cite{10.1063/1.1672831} \\
			ZnGeP$_2$  &10.318 &20.239& 0.981 & 0.258 & \cite{10.1063/1.1678930}\\
			ZnSnP$_2$  &10.679 &21.358 &1.00 &0.239  & \cite{https://doi.org/10.1002/pssb.19680290144} \\
			MgSiP$_2$  &10.811 &19.077 & 0.882 &0.292 & \cite{SPRINGTHORPE1969,Vaipolin1973}\footnote{$a$ and $c$ from Ref. \cite{SPRINGTHORPE1969}; 
            $u$ from Ref. \cite{Vaipolin1973}.}
		\end{tabular}
	\end{ruledtabular}
\end{table}

\end{appendices}



%

\end{document}